\documentclass[prb, showkeys]{revtex4}

\usepackage{graphicx}
\usepackage{dcolumn}
\usepackage{bm}
\usepackage[cp1250]{inputenc}
\usepackage{ifthen} 

\begin{document}

\newcommand{\bk}{\mathbf{k}}
\newcommand{\bq}{\mathbf{Q}}
\newcommand{\bQ}{\mathbf{Q}}
\newcommand{\mb}{\mathbf}
\newcommand{\hw}{\hbar\omega_C}
\newcommand{\eq}{\begin{equation}}
\newcommand{\eqx}{\end{equation}}
\newcommand{\eqn}{\begin{eqnarray}}
\newcommand{\eqnx}{\end{eqnarray}}

\preprint{APS/123-QED}

\title{Superconductivity in an almost localized Fermi liquid of quasiparticles with spin-dependent masses and effective field induced by electron correlations}

\author{Jan Kaczmarczyk}
\email{jan.kaczmarczyk@uj.edu.pl}
\author{Jozef Spałek}
 \email{ufspalek@if.uj.edu.pl}
\affiliation{
Marian Smoluchowski Institute of Physics, Jagiellonian University, \linebreak ul. Reymonta 4, 30-059 Kraków, Poland
}
\date{\today}

\begin{abstract}
Paired state of nonstandard quasiparticles is analyzed in detail in two model situations. Namely, we consider the Cooper-pair bound state and the condensed phase of an almost localized Fermi liquid (ALFL) composed of quasiparticles in a narrow-band with the spin-dependent masses (SDM) and an effective field, both introduced earlier and induced by strong electronic correlations. Each of these novel characteristics are calculated in a self-consistent manner. We analyze the bound states as a function of Cooper-pair momentum $|\bq|$ in applied magnetic field in the strongly Pauli limiting case (i.e. when the orbital effects of applied magnetic field are disregarded). The spin-direction dependence of the effective mass makes the quasiparticles comprising Cooper pair \textit{spin distinguishable in the quantum mechanical sense}, whereas the condensed gas of pairs may still be regarded as composed of identical entities. The \textit{Fulde-Ferrell-Larkin-Ovchinnikov} (FFLO) condensed phase of moving pairs is by far more robust in the applied field for the case with spin-dependent masses than in the situation with equal masses of quasiparticles. Relative stability of the \textit{Bardeen-Cooper-Schrieffer} (BCS) vs. FFLO phase is analyzed in detail on temperature - applied field plane. Although our calculations are carried out for a model situation, we can conclude that the spin-dependent masses should play an important role in stabilizing high-field low-temperature (HFLT) unconventional superconducting phases (FFLO being an instance) in systems such as CeCoIn$_5$, organic metals, and possibly others.
\end{abstract}


\maketitle

\section{\label{sec:intro}Introduction}

Unconventional superconductivity of heavy-fermion \cite{a} and organic metals \cite{b} is studied almost as intensively as that of high-temperature superconductors \cite{c}. While the $d$-wave symmetry of the superconducting gap is shared among all of the systems, that have a quasi two-dimensional electronic structure, the first two systems are easier to approach theoretically, as the corresponding normal state can be conceptually described as a Fermi liquid, albeit an \textit{almost localized Fermi liquid}, ALFL \cite{d}, whereas the normal state of a high-temperature superconductor is an almost-localized \textit{non-Fermi quantum liquid}, at least on the underdoped side \cite{e}. By ALFL we understand the electron liquid located on phase diagram close to the Mott or Mott-Hubbard threshold, composed of (nonstandard) quasiparticles. The nonstandard characteristics of the quasiparticles comprise: (i) large value of their effective mass $m^*$ which becomes divergent if the Mott transition has the character of a \textit{quantum critical point} \cite{f}, (ii) spin-direction dependence of the effective masses $m^* \equiv m_\sigma$ in the magnetically polarized state in the non-half filled band situation \cite{g}, and (iii) the appearance of the effective field $h_{cor}$ induced by electronic correlations which differs from the molecular field introduced in magnetism \cite{h}. Thus, it is of basic interest to include those novel features in the description of concrete physical phenomena, as they may mimic non-Fermi liquid features even though they represent only non-Landau corrections to the Landau Fermi liquid theory. The approach developed here is based on the \textit{Bardeen-Cooper-Schrieffer} (BCS) theory of superconductivity and closely associated with it the problem of a single Cooper pair, both treated without and with the spin-dependent masses, in the latter case also with inclusion of the effective field $h_{cor}$ calculated within a self-consistent scheme. In this manner, our approach represents a natural and simple extension of both Landau Fermi liquid and BCS theories to the systems with correlated electrons.

\textit{Spin-dependent masses} (SDM) of quasiparticles have been observed recently in the heavy-fermion superconductor CeCoIn$_5$ \cite{McCollam} and other systems \cite{Sheikin} by means of the de Haas-van Alphen oscillations in strong applied magnetic field.
These observations confirm the earlier theoretical prediction concerning the spin-dependent mass enhancement induced by inter-electron correlations in a single narrow-band \cite{g} and in the Anderson-lattice systems \cite{Spalek1, Spalek2}. This phenomenon has also been reinvestigated recently within the Periodic Anderson \cite{Bauer} and the Hubbard \cite{Bauer2} models. It appears that SDM should occur for either moderately or strongly-correlated systems with large on-site Coulomb repulsion $U$. This effect is particularly strong near the half filling $n \rightarrow 1$ in the narrow-band or for an almost integer valency of Ce$^{+4-n_f}$ ($n_f \rightarrow 1$) in the heavy-fermion systems. Furthermore, in the same system - CeCoIn$_5$, in which SDM were observed, a novel high-field low-temperature (HFLT) superconducting phase has been discovered \cite{CeCo2}. This phase was proposed to be either of the \textit{Fulde-Ferrell-Larkin-Ovchinnikov} (FFLO) nature \cite{Fulde, Larkin} or an unconventional superconducting phase coexisting with an antiferromagnetic order \cite{AForder, Sigrist}, or even a complex phase with three independent order parameters \cite{Aperis1}. Hence, its nature is still unclear. Therefore, it is important to consider a case with SDM and reexamine the BCS phase against formation of the FFLO phase. Such analysis can help in predicting the influence of SDM on the HFLT stability. To achieve these goals we consider here only a model situation to emphasize the role of nonstandard quasiparticle characteristics in altering the standard BCS or FFLO description. In particular, we show that the FFLO state is robust in the SDM case on the expense of the BCS state. The simple model reflects some of the qualitative features of the phase diagram observed for CeCoIn$_5$ \cite{CeCo2}, although our main purpose here is to underline the universal properties rather than to construct the detailed phase diagram for concrete systems. In part of our numerical analysis, we take the values of some parameters (e.g. the elementary cell volume) such as for CeCoIn$_5$ to illustrate that the results are in the proper range of temperature and applied field, although the detailed treatment for this system must be carried out separately and include also the d-wave symmetry, the singlet-triplet mixing in the paired state, and possibly, the spin-density-wave appearance, coexisting with the proposed here generalized FFLO state. We also discuss briefly some of the nontrivial features of the normal state.

The structure of the paper is as follows. In Sec. II we analyze the concept of electron gas with SDM and of the effective field, and introduce its basic characteristics in the normal state. In Sections III and IV we discuss a single Cooper-pair problem \cite{Cooper} for such Fermi liquid, particularly in applied magnetic field and obtain a stable pair state with center-of-mass momentum $\bq \neq 0$. The pair is discussed at some length, as it conveys a basic message concerning the formation of FFLO state and is nontrivial because quasiparticles composing it are spin distinguishable. In Sections V and VI we analyze the condensed state of pairs, both analytically and numerically. Finally, Sec. VII contains outlook and concluding remarks. In Appendices \ref{appA} - \ref{appC} we discuss the details concerning the origin of SDM, introduce the general antisymmetric state of Cooper pair, and justify the narrow-band limit for the case of Anderson lattice with an intraatomic hybridization, respectively.

\section{Quasiparticle gas with heavy spin-dependent masses and effective field induced by correlations \label{sec:Formulation}}

We analyze first the normal state properties of a quasiparticle gas with the spin-direction ($\sigma = \pm 1$) dependent masses $m^* \equiv m_\sigma$ and the effective field induced by correlations $h_{cor}$. Quasiparticle energies in the applied field $h \equiv g \mu_B H_a$ have the form

\begin{equation}
    \xi_{\bk \sigma} = \frac{\hbar^2 k^2}{2 m_\sigma} - \sigma h - \mu - \sigma h_{cor}, \label{eq:disp}
  \end{equation}

where we have taken the simple parabolic dispersion relation and have defined from the start the energy with respect to the chemical potential $\mu$. The spin dependence of the masses \cite{g, Spalek2} is taken in the simplest form corresponding to the narrow-band or the Kondo-lattice limits (cf. Appendices \ref{appA} and \ref{appC}) with the Hubbard interaction $U \rightarrow \infty$, i.e.

\begin{equation}
    \frac{m_\sigma}{m_B} = \frac{1-n_\sigma}{1-n} = \frac{1-n/2}{1-n} - \sigma \, \frac{\overline{m}}{2(1-n)} \equiv \frac{1}{m_B} (m_{av} - \sigma \Delta m/2),
    \label{eq:m}
\end{equation}

  where $\sigma = \pm1$ is the spin quantum number, $m_B$ is the band mass, $\overline m \equiv n_\uparrow - n_\downarrow$ is the system magnetic polarization and $n$ is the total band filling ($n = n_\uparrow + n_\downarrow)$. Also, $\Delta m \equiv m_2 - m_1$ is the mass difference and $m_{av} \equiv (m_1 + m_2) / 2$ is the average mass.
  Note that in the magnetic saturation limit $(n_\uparrow - n_\downarrow)/(n_\uparrow + n_\downarrow) =  1$ we recover the band limit with $m_\uparrow / m_B = 1$, whereas the heavy quasiparticles in the spin-minority disappear ($n_\downarrow = 0$). Note also that the convention is such that the state $\sigma = +1$ is regarded as that with magnetic moment along the applied field direction.

  In this paper we follow the spirit of original paper by Fulde and Ferrell \cite{Fulde} and adopt it to the ALFL case. In such formulation, the system of self-consistent equations determining thermodynamic properties of the normal state starting from the free-energy functional $\mathcal{F}$, is as follows

\begin{eqnarray}
&& \mathcal{F} = - k_B T \sum_{\bk \sigma} \ln(1+e^{-\beta \xi_{\bk \sigma}}) + \mu N + \frac{N}{n} \overline{m} h_{cor}, \label{eq:start1b} \\
&& h_{cor} = - \frac{n}{N} \sum_{\bk \sigma} f(\xi_{\bk \sigma}) \frac{\partial \xi_{\bk \sigma}}{\partial \overline{m}}, \label{eq:start2b}\\
&& \overline{m} = \frac{n}{N} \sum_{\bk \sigma} \sigma f(\xi_{\bk \sigma}), \label{eq:start3b}\\
&& n = n_\uparrow + n_\downarrow = \frac{n}{N} \sum_{\bk \sigma} f(\xi_{\bk \sigma}),\label{eq:start4b}
\end{eqnarray}

   where $f(\xi_{\bk \sigma})$ is the Fermi-Dirac distribution, $\beta = 1/(k_B T)$ is the inverse temperature, and $N$ is the total number of particles. The free energy functional $\mathcal{F}(T, H_a; h_{cor}, \overline{m}, n)$ given by (\ref{eq:start1b}) describes a Fermi sea with the spin-dependent masses $m_\sigma$ and correlation field $h_{cor}$. The equations (\ref{eq:start2b}) and (\ref{eq:start3b}) are derived from the conditions $\partial \mathcal{F} / \partial \overline{m} = 0$ and $\partial \mathcal{F} / \partial h_{cor} = 0$ respectively, and the last equation (\ref{eq:start4b}) is an elementary relation used to fix the band filling (defined by $n / V_{elem} = N / V$, where $V_{elem}$ is the elementary cell volume). The normal-state properties determined via Eqs. (\ref{eq:start1b}) - (\ref{eq:start4b}) are to be compared with those for the paired state obtained in the subsequent sections.

   The equations describing the Fermi sea characteristics can be easily solved numerically by their reduction to a single equation for $n_\sigma$ of the following form

\begin{equation}
\frac{n \, n_1^{2/3}}{ (n-n_1)(2m_{av}-m_B)+m_B n_1 } = \frac{n (n - n_1)^{2/3}}{ n_1 (2m_{av} -
m_B) + m_B (n - n_1)} + \frac{4 (h + h_{cor})}{\hbar^2} \, \Big(\frac{V}{6\pi^2}\Big)^{2/3},
\end{equation}

\begin{figure*}
\scalebox{1.2}{\includegraphics{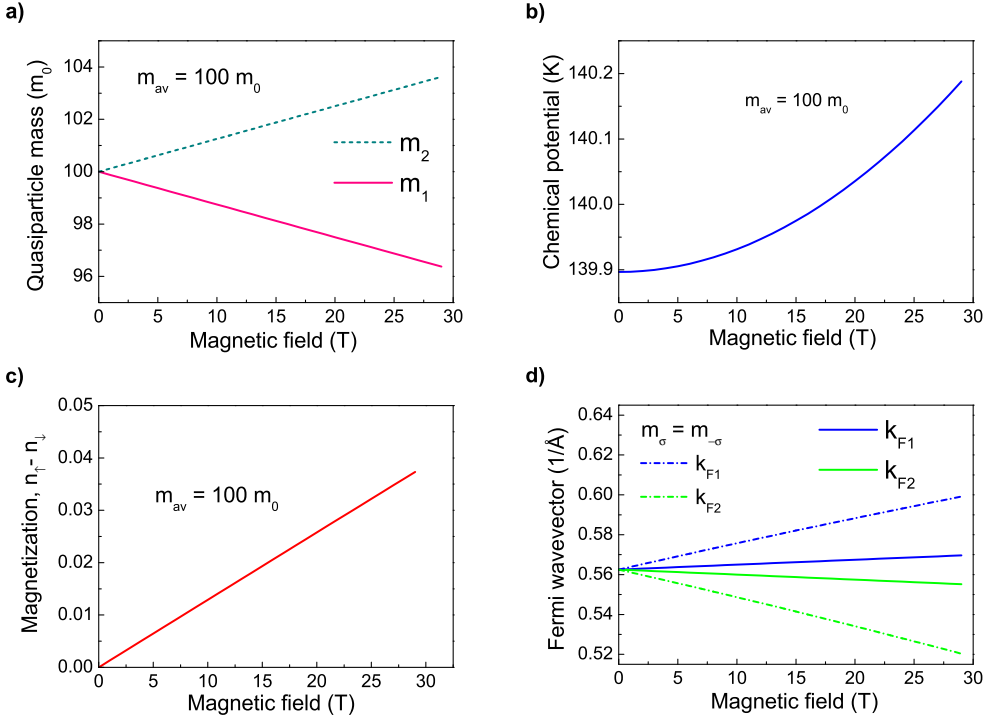}}
\caption{\label{fig:FS} (Color online) Panel with the Fermi sea characteristics in the normal state as a function of applied magnetic field (for $T = 0.05 \, K$). Dashed line in a) represents the mass in the spin-minority subband, whereas the solid line characterizes that in the spin-majority subband. The dotted-dashed lines in d) represent the results for subband Fermi wave vector in the case with spin-independent masses (SIM) with $m^* = m_{av}$. Note much greater Fermi wavevector splitting in the SIM case; this is important for understanding of the results for superconducting state. For details see main text.}
\end{figure*}

with $n_1 \equiv n_\uparrow$. The Fermi sea characteristics are summarized in Fig. \ref{fig:FS}. The mass difference, the Fermi vector splitting and magnetization increase linearly with the increasing field. Therefore, the approximated expressions in the field up to $30 \, T$ are $\overline{m} = \chi H_a$ and $m_\sigma(H_a) = m_{av} - \sigma m_B \chi / 2(1-n) \, H_a$. In our numerical study we obviously use full expressions (\ref{eq:start1b}) - (\ref{eq:start4b}). Namely, we take the values of parameters emulating the heavy-fermion systems: $n = 0.97$, $V_{elem} = 161 \AA^3$ and the $h=0$ value of the quasiparticle mass $m_{av} \equiv m_B \frac{1-n/2}{1-n} = 100 \, m_0$. For these values the principal characteristics are collected in Fig. \ref{fig:FS}. The assumed mass enhancement magnitude corresponds to the moderate heavy fermions, with the value of $\gamma$ in the range $100-200 \, mJ / mol \, K^2$. Also, the value of $n = 0.97$ corresponds to the effective valence of the Ce ions $+4-n_f = 3.03$, a typical value. Note that the mass splitting is only about $7\%$ in the field of $H_a = 30 \, T$, but more important is the Fermi wavevector splitting $\Delta k_F \equiv k_{F\uparrow} - k_{F\downarrow}$, displayed in Fig. \ref{fig:FS}d. The Fermi wavevector is calculated according to the relation $k_{F\sigma} = (6 \pi^2 n_\sigma / V_{elem})^{1/3}$. Most of the characteristics are indeed linear in $H_a$, as stated above.

\begin{figure*}
\scalebox{0.7}{\includegraphics{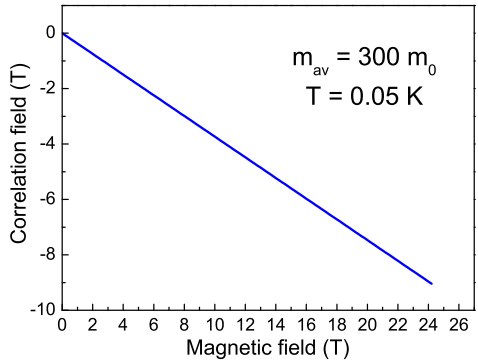}}
\caption{\label{fig:hcorFS} Correlation field $h_{cor}$ as a function of the applied magnetic field for the normal state. The linear dependence is $h_{cor} = -0.4 \, h$.}
\end{figure*}

In Fig. \ref{fig:hcorFS} we display the $h$ dependence of the effective field $h_{cor}$. It is linear in $h$ and typical values are $h_{cor} \approx -0.4 \, h$. More importantly, it is always antiparallel to the applied field, it partly compensates it in the sense that it reduces the Zeeman contribution to the quasiparticle energy. Also, the external field induces the effective-mass splitting and this factor drastically decreases the difference $\Delta k_F$.

\section{Cooper pair state: the quasiparticle distinguishability}
\label{sec:Cooper}

We consider next the simplest situation involving quasiparticle pairing - \textit{the Cooper problem} \cite{Cooper} and demonstrate principal and nontrivial features of the introduced quasiparticle states, particularly in the applied magnetic field. These features involve not only the pair binding energy, but raise also some basic questions concerning its quantum mechanical description, as discussed next.

\subsection{Hamiltonian with scalar effective masses and its limitations}

A simple way of accounting for the different masses of quasiparticles is to insert them into the Hamiltonian "by hand", that is to assume that the first particle has the spin quantum number $\sigma = \uparrow \, \equiv \, 1$, and the corresponding mass $m_\uparrow \equiv m_1$, whereas the other one has spin $\sigma = \downarrow \, \equiv \, -1$, and mass $m_\downarrow \equiv m_2$. This is a standard procedure, since in nonrelativistic quantum mechanics the particle mass is an external, classical parameter. This assumption leads to the following form of the Hamiltonian for the two particles with $m_\uparrow = m_1$ and $m_\downarrow = m_2$

\begin{equation}
\mathcal{H} = -\frac{\hbar^2}{2 m_1}\nabla_1^2 - \frac{\hbar^2}{2 m_2}\nabla_2^2 - \sigma_1^z h - \sigma_2^z h - \sigma_1^z h_{cor} - \sigma_2^z h_{cor} + V(\mathbf{r_1}, \mathbf{r_2}) \label{eq:Hscal}.
\end{equation}

Where $V(\mathbf{r_1}, \mathbf{r_2})$ is an attractive interaction between them. We neglect the effect of applied magnetic field on the electron orbital, because the Maki parameter \cite{Maki} is usually large in correlated systems (see e.g. CeCoIn$_5$, in which $\alpha^\parallel = 4.6$, $\alpha^\perp = 5.0$ \cite{CeCo2}), which means that the Pauli contribution dominates over the orbital effects. For the two-particle state with opposite spins, Hamiltonian (\ref{eq:Hscal}) acts properly only on the wave function with the spin part of the form

\begin{equation}
\chi(\mathbf{\sigma_1}, \mathbf{\sigma_2}) = |1\uparrow\rangle|2\downarrow\rangle,  \label{eq:spin1}
\end{equation}

This type of wave function will be called \textit{specific-spin} in the following. If we applied the Hamiltonian to the wave function with different spin part, i.e. $\chi'(\mathbf{\sigma_1}, \mathbf{\sigma_2}) = |1\downarrow\rangle|2\uparrow\rangle$, it would assign improperly mass $m_1 \equiv m_\uparrow$ to the first quasiparticle, whose spin is $\sigma = \downarrow$. For this reason, we may not construct for Hamiltonian (\ref{eq:Hscal}) the proper singlet or triplet spin wave functions $\chi_{S, T}(\mathbf{\sigma_1}, \mathbf{\sigma_2}) = \big( |1\uparrow\rangle |2\downarrow\rangle \mp |1\downarrow\rangle|2\uparrow\rangle \big) / \sqrt{2} $, in which the spin quantum number has not been assigned to the particle of given mass. This means that by specifying that the first particle has the spin up, and the second the spin down we violate in an obvious manner the quasiparticle-spin \textit{indistinguishability}.

One can illustrate the nontrivial character of the present case with SDM by commenting on the simplest quantum system - the hydrogen atom. Namely, although the masses of proton and electron are vastly different, the total spin part of the wave function is still the full singlet. This is because the spin quantum numbers of those particles are completely detached from their masses. In effect, the spin part of the wave function is either pure singlet or triplet. On the contrary, in our situation with SDM, the spin quantum number is attached to the masses, so the spin transposition symmetry may be broken explicitly, as discussed in detail below.

\subsection{Hamiltonian with masses in an invariant (operator) form}

We construct next the Hamiltonian in such a way, that it properly assigns masses to quasiparticles depending on their spin direction in the applied magnetic field. The only way to do this is to introduce the mass operator $\hat{m}(\sigma_i^z) \equiv m_{av} - \frac{1}{2} \sigma_i^z \, \Delta m$, where the mass splitting $\Delta m$ needs not to be specified at this point. Under this prescription, the two-particle Hamiltonian takes the form

\begin{widetext}
\begin{equation}
\mathcal{H} = -\frac{\hbar^2}{2 m_{av} - \sigma_1^z \, \Delta m}\nabla_1^2 - \frac{\hbar^2}{2 m_{av} - \sigma_2^z \, \Delta m}\nabla_2^2 - \sigma_1^z h - \sigma_2^z h - \sigma_1^z h_{cor} - \sigma_2^z h_{cor} + V(\mathbf{r_1}, \mathbf{r_2}) \label{eq:HMOP}.
\end{equation}
\end{widetext}

Now, the kinetic part assigns respective masses $m_1 = m_{av} - \Delta m / 2$ and $m_2 = m_{av} + \Delta m / 2$ to the particles depending on their spin z-component. With this Hamiltonian we can now analyze any spin function. In particular, both the specific-spin wave function (\ref{eq:spin1}), as well as those describing the singlet or triplet states with the $z$-component of the total spin $S_1^z + S_2^z = 0$, which have obviously the form

\begin{eqnarray}
\chi_{S, T}(\sigma_1, \sigma_2) = \frac{1}{\sqrt{2}} \Big( |1\uparrow\rangle|2\downarrow\rangle \mp |1\downarrow\rangle|2\uparrow\rangle \Big).
\label{eq:spin2}
\end{eqnarray}

Additional advantage of the Hamiltonian (\ref{eq:HMOP}) is the ability to describe transition from the \textit{indistinguishable-quasiparticles} limit (for zero or low mass splitting) to the \textit{spin-distinguishable-quasiparticles} case when $h > 0$. By distinguishable quasiparticles we mean here those described by a wave function without well-defined transposition symmetry (i.e. non-antisymmetric in our case).

\subsection{Singlet, triplet, and their inadequacy in describing the paired state}

First, we note that the total spin operator $\hat{\mathbf{S}}^2 \equiv (\hat{\mathbf{S_1}} + \hat{\mathbf{S_2}})^2$ does not commute with Hamiltonian (\ref{eq:HMOP}), i.e.

\begin{equation}
  [\mathcal{H}, (\hat{\mathbf{S_1}} + \hat{\mathbf{S_2}})^2] \neq 0, \label{eq:commute}
\end{equation}

whereas $\mathcal{H}$ does commute with $\hat{S_1^z}$ and $\hat{S_2^z}$ separately. This means that while the z-components of the \textit{individual spins} represent good quantum numbers, the \textit{total length does not}. To analyze this property, it is important to see that the spin-dependent denominators can be rewritten in the two equivalent forms

\eqn
\mathcal{H} & = & \sum_{i=1}^2 \Bigg[ - \frac{\hbar^2}{2 m_{av} - \sigma_i^z \Delta m} \nabla_i^2 -  \sigma_i^z h -  \sigma_i^z h_{cor} \Bigg] + V(\mathbf{r_1}, \mathbf{r_2}) = \nonumber \\
& \equiv & \sum_{i = 1}^2 \Bigg[ - \frac{\hbar^2}{4 m_{av}^2  - (\Delta m)^2} (2 m_{av} + \sigma_i^z \Delta m ) \nabla_i^2 -  \sigma_i^z h -  \sigma_i^z h_{cor} \Bigg] + V(\mathbf{r_1}, \mathbf{r_2}).
\eqnx

Taking $\hat{\mathbf{S_i}} = (1/2) \hat{\mathbf{\sigma_i}}$, one can easily prove the condition (\ref{eq:commute}), as well as the property that $[\mathcal{H}, \hat{S_i^z}] = 0$. One sees that the spin wave function can be characterized by individual values $\sigma_1^z = \uparrow$ and $\sigma_2^z = \downarrow$ or vice versa, \textit{but the two-particle spin state might not have a proper singlet or triplet symmetry} (\ref{eq:spin2}). Note also that the property (\ref{eq:commute}) is independent of the form of pairing potential and is fulfilled also for $V(\mathbf{r_1}, \mathbf{r_2}) = 0$. This may lead also to the normal-state corrections not discussed here.

\subsection{Solution for the specific-spin wave function}

The wave function is decomposed into the spin and space parts

\begin{equation}
 \Psi(\mathbf{r_1}, \mathbf{r_2}, \sigma_1, \sigma_2) = \Phi( \mathbf{r_1}, \mathbf{r_2} ) \chi(\mathbf{\sigma_1}, \mathbf{\sigma_2}), \label{eq:sep}
\end{equation}

with the space part given as a superposition of plane wave states, i.e.

\begin{widetext}
\begin{equation}
\Phi( \mathbf{r_1}, \mathbf{r_2} ) = \sum_{\mathbf{k_1}, \mathbf{k_2}} \alpha_{\mathbf{k_1},
\mathbf{k_2}} \, \Psi_\mathbf{k_1}(\mathbf{r_1}) \Psi_\mathbf{k_2}(\mathbf{r_2}) = \frac{1}{V}
\sum_{\mathbf{k_1}, \mathbf{k_2}} \alpha_{\mathbf{k_1}, \mathbf{k_2}} \, e^{i \mathbf{k_1 r_1} + i
\mathbf{k_2 r_2}}.\label{eq:psi0}
\end{equation}
\end{widetext}

By starting from the wave function in the form (\ref{eq:sep}) with the space part given by (\ref{eq:psi0}) and spin part $\chi(\mathbf{\sigma_1}, \mathbf{\sigma_2}) = |1\uparrow\rangle|2\downarrow\rangle$, we can solve the Schr\"{o}dinger equation (with any of the two Hamiltonians) in a similar manner as in the original Cooper problem \cite{Cooper}. One of the differences we introduce is a new definition of the relative momentum when transforming the Hamiltonian to the center-of-mass and relative coordinates, namely

\begin{eqnarray}
&&\mathbf{R} =  \frac{\mathbf{r_1} m_1 + \mathbf{r_2} m_2}{m_1 + m_2};  \quad \bQ  =
\mathbf{k_1} + \mathbf{k_2}, \label{eq:trafom1}\\
&&\mathbf{r}  =  \mathbf{r_1} - \mathbf{r_2}; \quad \,\,\, \mathbf{k}  =  \frac{\mathbf{k_1} m_2 -
\mathbf{k_2} m_1}{m_1 + m_2}, \label{eq:trafom2}
\end{eqnarray}

where $\bQ$ is the total momentum and $\bk$ is called, by analogy to the standard Cooper problem, the relative momentum \cite{Wrobel}. After this transformation the Hamiltonian and wave function can be cast to the forms

\begin{eqnarray}
&&\mathcal{H} = - \frac{\hbar^2}{2 M}\nabla_\mathbf{R}^2 -\frac{\hbar^2}{2 \mu_{red}}\nabla_\mathbf{r}^2 - \sigma_1^z h - \sigma_2^z h - \sigma_1^z h_{cor} - \sigma_2^z h_{cor} + V(r), \label{eq:HCM2}\\
&&\Psi(\mathbf{R}, \mathbf{r}, \sigma_1, \sigma_2) = \frac{1}{V} e^{i \bQ \mathbf{R}} \sum_{\mathbf{k}} \alpha_{\mathbf{k}} \, e^{i \mathbf{k r}} |1\uparrow\rangle|2\downarrow\rangle,   \label{eq:psi3}
\end{eqnarray}

where $M = m_1 + m_2$ and $\mu_{red} = m_1 m_2/M$. Following the standard procedure, we obtain the equation determining coefficients $\alpha_\bk$ and the eigenenergy $E$ in the form

\begin{equation}
\alpha_\mathbf{k} = - \frac{1}{N} \frac{\sum_\mathbf{k'} V_{\mathbf{k k'}} \alpha_\mathbf{k'}}
{\epsilon_\bQ + \epsilon_\mathbf{k} - E} \label{eq:coeffm1},
\end{equation}

where $\epsilon_\bQ \equiv \hbar^2 \bQ^2 / 2 M$, $\epsilon_\bk \equiv \hbar^2 \bk^2 / 2 \mu_{red}$ and $N$ is the total number of particles. The interaction region has to be defined in a more general way, since we want to describe a system with nonequal Fermi vectors and possibly, with a non-zero center-of-mass momentum $\bQ$, which is a constant of motion. We assume that a constant, attractive interaction takes place in the regions of $\bk$-space for which both particles are at most at the distance $\hbar \omega_C$ above their Fermi surface, i.e.

\begin{equation}
V_{\mathbf{k k'}} = \left\{
\begin{array}{ll} -V_0, &\textrm{for } \xi_{\bk_1 \uparrow}, \xi_{\bk_2 \downarrow}, \xi_{\bk'_1 \uparrow}, \xi_{\bk'_2 \downarrow} \in [0, \hbar \omega_C],\\
0, & \textrm{in other cases,}
\end{array} \right.
\label{eq:potm1}
\end{equation}

\begin{figure}
\scalebox{0.22}{\includegraphics{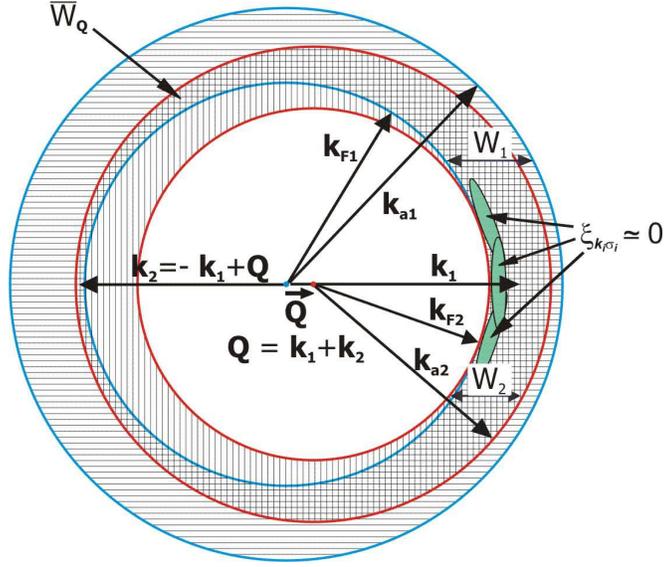}}
\caption{\label{fig:W} (Color online) Interaction region $\overline{W}_\bQ$ is the common part of $W_1$ and $W_2 + \bQ$ shifted by $- \frac{m_1}{M} \bQ$. The vectors $\mathbf{k_{ai}}$, are defined by $\xi_{\mathbf{k_{a1}} \uparrow} = \xi_{\mathbf{k_{a2}} \downarrow} = \hbar \omega_C$. Regions which contribute most to the pairing are marked as the three (green) ovals on the right with $\xi_{\bk_i \sigma_i} \approx 0$. For details see main text.}
\end{figure}

where $\bk$ and $\bQ$ are related to $\bk_1$, and $\bk_2$ via transformation (\ref{eq:trafom1}) and (\ref{eq:trafom2}) (the same holds for the vectors pairs ${\bk', \bQ}$ and ${\bk'_1, \bk'_2}$). We call $\overline{W}_\bQ$ the region in $\bk$-space in which the interaction is nonzero (cf. Fig. \ref{fig:W}). It can be shown that $\overline{W}_\bQ = [W_1 \cap \big(W_2 + \bQ\big)] - \frac{m_1}{M} \bQ$, where by adding a vector to the region in $\bk$-space we mean the whole region shifted by that vector. Also, $W_i = \{ \mathbf{k} \,|\, 0 \leq \xi_{\bk\sigma_i} \leq \hbar \omega_C \} \textrm{, } i=1,2$. In this notation, equation for the binding energy $\Delta \equiv 2 \mu - E$  becomes

\begin{widetext}
\begin{equation}
\frac{N}{V_0} = \frac{V}{8 \pi^3}
\int_{\overline{W}_\bQ} \frac{d^3 k} {\epsilon_\mathbf{\bQ} + \epsilon_\mathbf{\bk} - 2 \mu + \Delta} \equiv \frac{V}{8 \pi^3}
\int_{\overline{W}_\bQ} \frac{d^3 k} {\xi_\mathbf{k_1\uparrow} + \xi_\mathbf{k_2\downarrow} + \Delta}, \label{eq:gapC3b}
\end{equation}
\end{widetext}

From this form of the equation for $\Delta$ we can deduce that the regions of reciprocal space contributing most to the pairing are those for which $\xi_\mathbf{k_i\sigma_i} \simeq 0$ (see Fig. \ref{fig:W}). Large part of the space fulfills this condition if $|\bQ| \simeq \Delta k_F \equiv k_{F1} - k_{F2}$. Therefore, we can anticipate that the pair will have maximum binding energy when the pair center-of-mass momentum is close to the Fermi vector splitting, i.e. when $|\bQ| \simeq \Delta k_F$, consistent with the results for the FFLO state.

The equation (\ref{eq:gapC3b}) for the gap $\Delta$ has to be solved numerically for each $|\bQ|$ and the final solution in our approximation is the one with the largest binding energy. In the case of Cooper pair at rest ($\bQ = 0$), equation (\ref{eq:gapC3b}) can be solved analytically. Such analytic solution depends on which of the two vectors $\mathbf{k_{ai}} = \frac{1}{\hbar} \sqrt{2m_i(\mu + \sigma_i h + \sigma_i h_{cor} + \hbar \omega_C)}$ is larger, namely

\begin{widetext}
\begin{equation}
    \Delta = \left\{
    \begin{array}{ll}
    \frac{\hbar \omega_C \frac{M}{m_1}}{e^{C} - 1} \,
- \, \big[ (h+h_{cor}) M - \mu \Delta m \big] \frac{1}{e^{C} - 1}  \Big(\frac{1}{m_1} + \frac{1}{m_2} \, e^{C} \Big), & \textrm{for } k_{a1} \geq k_{a2},\\
   \frac{\hbar \omega_C \frac{M}{m_2}}{e^{C} - 1} - \frac{1}{m_2}\big[ (h + h_{cor}) M - \mu \Delta m \big], & \textrm{for } k_{a1} < k_{a2},
    \end{array}
    \right.  \label{eq:DelAnalGen}
\end{equation}
\end{widetext}

where $C \equiv \frac{2\pi^2 \hbar^2 N}{V \, V_0 k_{F1} \mu_{red}}$. In zero magnetic field we have that $C = \frac{2}{V_0 \overline{\rho}(\epsilon_F)}$ and expression (\ref{eq:DelAnalGen}) properly reproduces the Cooper result $\Delta = 2\hbar\omega_C/\big[\exp(\frac{2}{V_0 \overline{\rho}(\epsilon_F)})-1\big]$, where $\overline{\rho}(\epsilon_F)$ is the average density of states per particle.

\section{Detailed analysis of the Cooper-pair state}

\subsection{\label{sec:singlet}Is the singlet state a proper eigenstate?}

As has already been said, Hamiltonian (\ref{eq:HMOP}) does not commute with the total spin $\hat{\mathbf{S}}^2$. This means that the singlet and the triplet states are not a good basis for the problem considered. Nevertheless, we can still analyze energy for the singlet spin function (and we can think of it as a superposition of eigenfunctions with spin parts $|1\uparrow\rangle|2\downarrow\rangle$ and $|1\downarrow\rangle|2\uparrow\rangle$). This means, we can take a particular solution in the form

\begin{widetext}
\begin{equation}
\Psi(\mathbf{r_1}, \mathbf{r_2}, \sigma_1, \sigma_2) = \frac{1}{V}
\sum_{\mathbf{k_1}, \mathbf{k_2}} \alpha_{\mathbf{k_1}, \mathbf{k_2}} \, e^{i \mathbf{k_1 r_1} + i
\mathbf{k_2 r_2}} \,\, \frac{1}{\sqrt{2}} \big( |1\uparrow\rangle|2\downarrow\rangle - |1\downarrow\rangle|2\uparrow\rangle \big). \label{eq:PsiSinglet}
\end{equation}
\end{widetext}

Solution of the Schr\"{o}dinger equation in this case is quite cumbersome and will not be presented in detail here. It can be shown that, for the resulting equations not to be contradicting, the interaction has to be introduced as nonzero in the region (for derivation see Appendix \ref{appB})

\begin{widetext}
\begin{equation}
W^* = \{ \mathbf{k} = \frac{\mathbf{k_1} m_2 - \mathbf{k_2} m_1}{m_1 + m_2}
\,|\, \mathbf{k_1} \in W_1 \cap W_2 \wedge \mathbf{k_2} \in W_2 \cap W_1 \wedge \bQ = \mathbf{k_1} + \mathbf{k_2} \},
\end{equation}
\end{widetext}

or equivalently,

\begin{equation}
  W^* = \overline{W}_\bQ \cap (- \overline{W}_\bQ + \bQ\frac{\Delta m}{M}).
\end{equation}

The replacement of $\overline{W}_\bQ$ with $W^*$ presents the only difference in the case of pure singlet state, with respect to the that for the \textit{specific-spin state} $|1\uparrow\rangle|2\downarrow\rangle$ or $|1\downarrow\rangle|2\uparrow\rangle$. This interaction region coincides with $\overline{W}_\bQ$ if and only if $\bQ = 0$. As a result, the energy of the singlet state in the case of Cooper pair at rest is identical to the energy for the specific-spin states. For a moving pair ($\bQ \neq 0$) the interaction region is reduced, and the binding energy for the singlet state decreases rapidly with increasing $|\bQ|$, whereas that for the specific-spin states can become even larger (see Fig. \ref{fig:GBQ}).

\subsection{\label{sec:sym}Quasiparticle distinguishability}

The wave function of fermions has to be antisymmetric with respect to transposition of spin and space coordinates

\begin{equation}
(\mathbf{r_1}, \sigma_1) \leftrightarrow (\mathbf{r_2}, \sigma_2). \label{eq:trafocz}
\end{equation}

For the Cooper pair this implies, that

\begin{equation}
\Psi(\mathbf{r_1}, \mathbf{r_2}, \sigma_1, \sigma_2) = - \Psi(\mathbf{r_2}, \mathbf{r_1}, \sigma_2, \sigma_1). \label{eq:sym}
\end{equation}

It turns out that the Cooper pair at rest (with the singlet spin part) has proper symmetry even for $m_1 \neq m_2$, a very interesting result. As long as $\bQ = 0$ leads to the minimum energy state, the wave function is antisymmetric and describes \textit{indistinguishable quasiparticles}.

\begin{figure}
\includegraphics[width=0.5\textwidth]{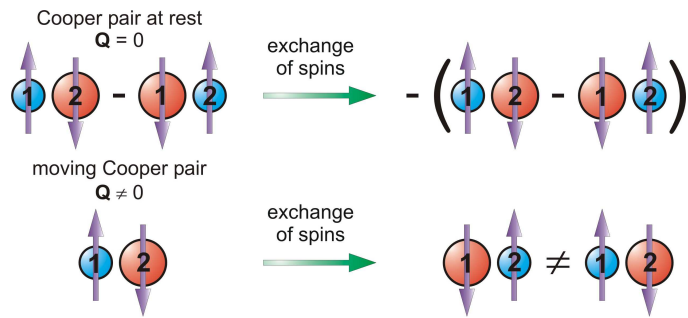}
\caption{\label{fig:kulki}(Color online) Behavior of the spin part of the wave function under transposition of particles labeled 1 and 2 for Cooper pair at rest (upper row, antisymmetry) and moving Cooper pair (lower row, no definite symmetry). }
\end{figure}

For sufficiently strong applied magnetic field the state with $\bQ \neq 0$ becomes stable and the pair has to be either in the $|1\uparrow\rangle |2\downarrow\rangle$ or $|1\downarrow\rangle |2\uparrow\rangle$ state. Such wave function does not have the symmetry (\ref{eq:sym}) and the transition to \textit{spin-distinguishable quasiparticles} takes place. The difference between the two situations has been illustrated in Fig. \ref{fig:kulki} and \ref{fig:GBQ}, the former providing the connection of the wave-function symmetry with the type of the solution.

A word of explanation is most proper at this point. Namely, the microscopic many-body Hamiltonian (Hubbard or Anderson-lattice, for example) respects the particle indistinguishability. This microscopic model is approximated here by an effective quasiparticle picture with the SDM \cite{g, Spalek2, Bauer, Bauer2}. The effective Hamiltonian taking into account those masses violates the indistinguishability in an obvious manner (see relation (\ref{eq:commute})). However, this quasiparticle picture has been confirmed experimentally \cite{McCollam, Sheikin} for the normal state. Therefore, this fact suggests that there is a basic \textit{qualitative} difference between the \textit{Landau Fermi liquid} describing rather weakly interacting fermions and the \textit{almost localized (local) Fermi liquid} more appropriate for moderately or strongly correlated particles.

\subsection{Numerical results}

In numerical analysis of the Cooper pair state we assume the following values of parameters, emulating the heavy-fermion systems: $n = 0.97, \,\, V_{elem} = 161 \, \AA^3$, $m_{av} \equiv m_B \frac{1-n/2}{1-n} = 100 \, m_0$ (all taken for CeCoIn$_5$). Also, we take $\hbar \omega_C = 100 \, K$ and $V_0 = 93 \, K$. Additionally, for sizeably different values of $\hbar \omega_C$, another maximum in the binding energy appears (between $\bQ = 0$ and $|\bQ| = \Delta k_F$), with binding energy maximally 5 mK higher than that for $\bQ = 0$. Therefore, with this third solution no new physics is incorporated in our study, and it might only blur the image presented. For the assumed parameters, the chemical potential is $\mu \approx 140 K$, i.e. of the same magnitude as $\hbar \omega_C$ and $V_0$. The relative value of the parameters is of no primary importance here, since the solution is of nonperturbational nature.

\begin{figure*}
\scalebox{1.5}{\includegraphics{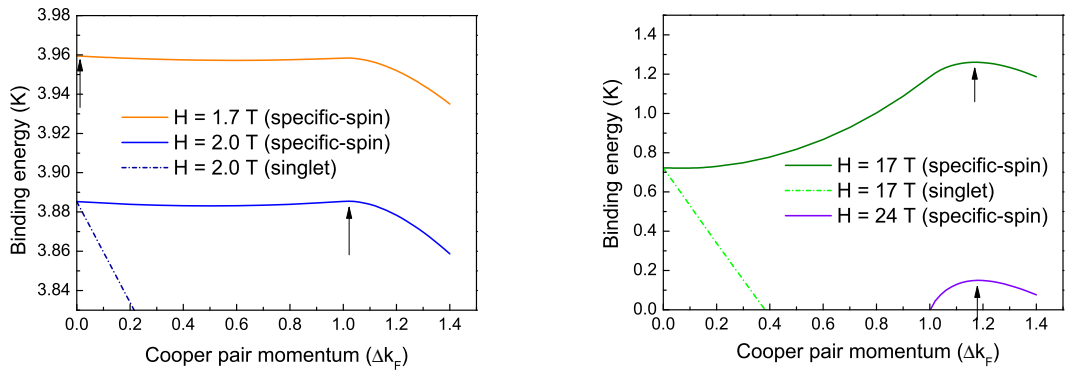}}
\caption{\label{fig:GBQ} (Color online) Binding energy of the specific-spin (solid lines) and singlet states (dot-dashed lines) as a function of center-of-mass momentum $\bQ$ and for selected values of applied field. Note a rapid decrease of binding energy of the singlet state. The bound state with $|\bQ| \neq 0$ becomes stable above the field $H_a \simeq 2 \, T$. The arrows mark the state with maximum value of binding energy.}
\end{figure*}

Solution of the gap equation (\ref{eq:gapC3b}) provides us with dependence of $\Delta$ on the Cooper pair center-of-mass momentum $\bQ$, in an applied field $H_a$, as shown in Fig. \ref{fig:GBQ}. For the  fields above $2 \, T$ the maximum binding energy appears for $|\bQ| \simeq \Delta k_F$. Note that the full singlet configuration (dot-dashed lines) has much smaller binding energy than the specific-spin state (solid lines).

\begin{figure}
\scalebox{1}{\includegraphics{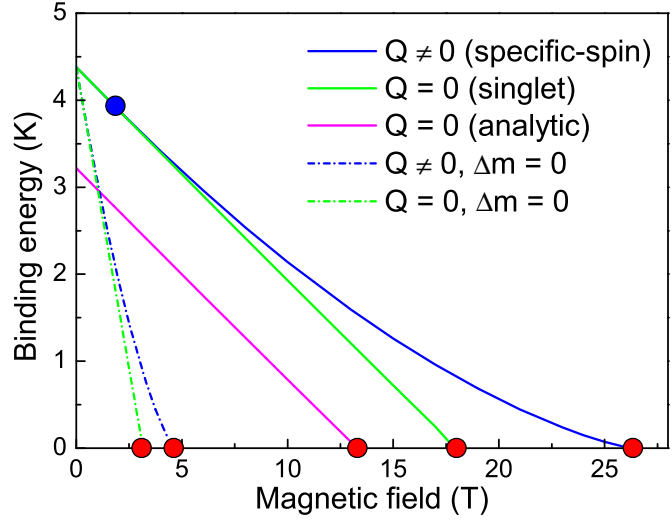}}
\caption{\label{fig:Gap} (Color online) Pair binding energy as a function of magnetic field for the cases with the spin-dependent (solid lines) and the spin-independent (dot-dashed lines) masses of quasiparticles. Note higher critical fields (as marked by the solid circles on the $x$-axis) for pair breaking in the former case. Analytic solution reproduces properly the dependence for $\bQ = 0$ state. The uppermost solid circle marks the transition from $\bQ = 0$ to $|\bQ| \simeq \Delta k_F$. The state with $\Delta m = 0$ and $\bQ \neq 0$ is nonphysical and is presented for comparison.}
\end{figure}

The binding energy as a function of the field is plotted in Fig. \ref{fig:Gap}. The case with SDM gives rise to much higher critical fields above which the pair is destabilized by the Pauli effect. The reason behind this robustness of the state for $\Delta m \neq 0$ is the smaller Fermi wavevector splitting $\Delta k_F$ for this case (c.f. Fig. \ref{fig:FS}), which in turn leads to the larger interaction region $\overline{W}_\bQ$. One can say, that the effect of SDM acts in the opposite direction than the Zeeman-term influence and compensates this influence to a degree.

\begin{figure}
\scalebox{0.16}{\includegraphics{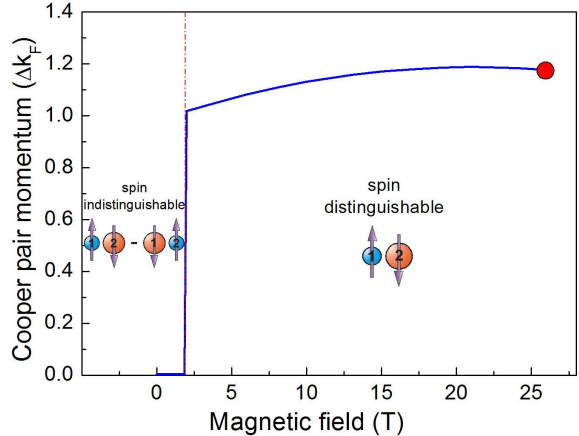}}
\caption{\label{fig:Qopt} (Color online) Optimal Cooper pair center-of-mass momentum versus magnetic field. For high enough fields $|\bQ| \simeq \Delta k_F$. The regimes with spin-singlet and specific-spin wave functions are also shown. The solid circle at the end marks the Pauli limiting critical field.}
\end{figure}

The optimal momentum value $|\bQ|$ vs. $H_a$ is displayed in Fig. \ref{fig:Qopt}. One observes a discontinuous change of $|\bQ|$ above the field $\simeq 2 \, T$ from $|\bQ| = 0$ to $|\bQ| \simeq \Delta k_F$. Above this point the specific-spin wave function provides the stable solution. At the same time, we have a transformation from the \textit{spin-indistinguishable} to \textit{spin-distinguishable} quasiparticles. In this manner, we have a model situation in which the question of distinguishability of quasiparticles can be investigated.

\subsection{An overview of the Cooper pair problem}
\label{sec5}

We have presented solution of the Cooper pair problem with SDM. The Hamiltonian has been formulated in two ways: with the \textit{scalar} masses and with \textit{operator} masses, the latter being more appropriate for the study of the two-particle pairing. The latter Hamiltonian also allows for systematic studies of the question of \textit{quasiparticle-spin distinguishability}. The resulting stable solution has an antisymmetric wave function and describes spin-singlet state for Cooper pair at rest ($\bQ = 0$), even if $m_1 \neq m_2$. However, for (and only for) $\bQ = 0$ the singlet solution is energetically degenerate with the specific-spin solution. For higher magnetic field, the solution with $\bQ = 0$ becomes energetically unfavorable, and this is why the transition to a moving-Cooper pair takes place. The moving Cooper-pair wave function has no definite transposition symmetry in the spin coordinates. Therefore, this transition is a typical transition to a broken symmetry state in the Landau sense, since the starting Hamiltonian with the effective masses in the operator form (cf. Eq.(\ref{eq:HMOP})) is written in spin-symmetric form. With the transition to a moving Cooper-pair state there also appears that from \textit{indistinguishable quasiparticles} to their \textit{spin-distinguishable correspondants}. This also leads to the admixture of the triplet component with $S^z = 0$ to the singlet state for $\bQ \neq 0$, as the \textit{specific-spin} function is a superposition: $|1\uparrow\rangle|2\downarrow\rangle = (|singlet\rangle + |triplet\rangle)/\sqrt{2}$. We think that the quasiparticle-spin distinguishability is thus indispensable, as the quasiparticles have different masses which represent their external characteristic, with the value of spin quantum number labeling them.

\section{Paired state of quasiparticles in almost localized Fermi liquid}

\subsection{General remarks on the pairing nature}

At present, two principal non-phonon mechanisms of pairing have been applied to strongly correlated electron systems. The first of them is the real-space pairing based on either $t$-$J$ model and applied to high temperature superconductivity \cite{1s} or on hybrid pairing applied to either heavy-fermion \cite{2s} or high-Tc superconductivity \cite{3s}. The second is based on (renormalized) paramagnon exchange \cite{4s}. In both cases, the effective pairing potential is explicitly momentum-dependent and this situation results in $\bk$-dependent gap. For the purpose of the present paper, in which we take the narrow-band limit of the Anderson lattice model with the $\bk$-independent hybridization matrix element (cf. Appendix \ref{appC}), the pairing potential near the Fermi level is approximately constant and thus leads to the $s$-wave form of the superconducting gap. Note however, that the original purpose of the present work is to formulate the language of description of the systems with SDM and $h_{cor}$ present in a simple situation. Such straightforward formulation leads already to quite nontrivial and probably universal mechanism of the HFLT phase stabilization by SDM (FFLO being an instance).

\subsection{Formulation of the problem}

We discuss next the condensed state of the nonstandard quasiparticles introduced in Sec. \ref{sec:Formulation}. In the paired state with isotropic gap the applied field penetrates only to the London depth $\lambda_L$, i.e. produces an irrelevant contribution of the spin-split-mass states onto bulk properties at temperature $T=0$. However, at $T>0$ the excited quasiparticles will produce a nonzero moment which should be clearly visible when the system approaches the transition temperature $T_s$. Moreover, SDM should provide an important contribution to the FFLO superconducting state, since in that phase there are normal regions. Therefore, in the situation with SDM (and with $h_{cor}$ reducing $h$) we should observe a robust FFLO phase in high fields on expense of the BCS state, as SDM help stabilizing it (by reducing the Fermi vectors splitting $\Delta k_F$ and by increasing the critical field due to the presence of $h_{cor}$). This is the subject of discussion in the next two Sections, as a result of which we construct the phase diagram on the applied field $H_a$ - temperature $T$ plane.

We introduce the BCS Hamiltonian with a constant pairing potential of magnitude $V_0$ (for its simple justification see Appendix \ref{appC}) and allow for the possibility of a nonzero center-of-mass momentum $\bq$ of a Cooper pair written in the representation symmetric with respect to both component quasiparticles composing the Cooper pair \cite{Shimahara}

\eq
\mathcal{H} = \sum_{\bk \sigma} \xi_{\bk \sigma} a^\dagger_{\bk \sigma} a_{\bk \sigma} - \frac{V_0}{N} \sum_{\bk \bk' \bq} a^\dagger_{\bk+\bq/2 \uparrow} a^\dagger_{-\bk + \bq/2 \downarrow} a_{-\bk' + \bq/2 \downarrow} a_{\bk' + \bq/2 \uparrow} + \frac{N}{n} \overline{m} h_{cor}, \label{eq:HStart}
\eqx

with $\xi_{\bk \sigma}$ given by (\ref{eq:disp}) taking account of the different quasiparticle masses. The magnetic field is accounted for only via the Zeeman term, as the Maki parameter \cite{Maki} in the systems of interest is high \cite{CeCo2} (Pauli limiting case). The interaction is assumed to exist only in region $\pm\hbar\omega_C$ around the Fermi surface, more precisely in the region

\eq
\overline{W} = \Big[\frac{k_{b\uparrow}+k_{b\downarrow}}{2}, \frac{k_{a\uparrow}+k_{a\downarrow}}{2}\Big],
\eqx

where $k_{b\sigma}$ is defined by $\xi_{k_{b\sigma}, \sigma} = -\hw$, and $k_{a\sigma}$, by $\xi_{k_{a\sigma}, \sigma} = \hw$. Such interaction region has been chosen because its width does not change significantly with the magnetic field $h$. We performed also calculations by selecting the interaction regime differently (namely, by choosing $\overline{W} = [k_{b\downarrow}, k_{a\uparrow}]$ and $\overline{W} = [k_{b\uparrow}, k_{a\downarrow}]$) and have obtained almost the same results. Hamiltonian (\ref{eq:HStart}) is diagonalized with standard mean-field procedure followed by the Bogolyubov-de Gennes transformation of the form

\eqn
&& \alpha_{\bk \uparrow} = u_\bk a_{\bk + \bq/2 \uparrow} - v_\bk a^\dagger_{-\bk + \bq/2 \downarrow}, \label{eq:Bog1}\\
&& \alpha^\dagger_{\bk \downarrow} = v_\bk a_{\bk + \bq/2 \uparrow} + u_\bk a^\dagger_{-\bk + \bq/2 \downarrow}, \label{eq:Bog2}
\eqnx

which leads to the diagonal form

\eq
\mathcal{H} = \sum_{\bk \sigma} E_{\bk \sigma} \alpha^\dagger_{\bk \sigma} \alpha_{\bk \sigma} + \sum_\bk (\xi^{(s)}_\bk - E_\bk) + N \frac{\Delta_\bq^2}{V_0} + \frac{N}{n} \overline{m} h_{cor},
\eqx

and the quasiparticle spectrum characterized by energies \cite{Shimahara}

\eqn
E_{\bk \sigma} & = & E_\bk + \sigma \xi^{(a)}_\bk, \quad \quad \quad \quad \quad \quad E_\bk = \sqrt{\xi^{(s)2}_\bk+\Delta_\bq^2}, \\
 \xi^{(s)}_\bk & \equiv & \frac{1}{2} (\xi_{\bk + \bq/2 \uparrow} + \xi_{-\bk + \bq/2 \downarrow}), \quad  \xi^{(a)}_\bk \equiv \frac{1}{2} (\xi_{\bk + \bq/2 \uparrow} - \xi_{-\bk + \bq/2 \downarrow}),
\eqnx

and with

\eq
 \Delta_\bQ  \equiv  \frac{1}{N} \sum_{\bk} \langle a_{-\bk + \bq/2 \downarrow} a_{\bk + \bq/2 \uparrow} \rangle
\eqx

being the $\bq$-dependent gap parameter. In the form (\ref{eq:Bog1})-(\ref{eq:Bog2}) of the Bogolyubov-de Gennes transformation the quasiparticle operators $\alpha_{\bk \uparrow}$ and $\alpha^\dagger_{\bk \downarrow}$ are distinguished by the pseudospin label $\uparrow$ and $\downarrow$. Note also that because of the presence of $\xi_\bk^{(a)}$, there are regions of reciprocal space for which $E_{\bk\sigma} \leq 0$, which represents nongapped excitations. The gap parameter $\Delta_\bq$ is determined from the self-consistent gap equation

\eq
\Delta_\bq = \frac{V_0}{N} \sum_\bk \frac{1 - f(E_{\bk \uparrow}) - f(E_{\bk \downarrow})}{2 E_\bk} \Delta_\bq.
\eqx

The Bogolyubov transformation coherence factors in (\ref{eq:Bog1}) and (\ref{eq:Bog2}) are given by

\eq
u_\bk = \Bigg[\frac{1}{2}\Big(1 + \frac{\xi^{(s)}_\bk}{E_\bk} \Big)\Bigg]^{1/2}, \quad
v_\bk = \Bigg[\frac{1}{2}\Big(1 - \frac{\xi^{(s)}_\bk}{E_\bk} \Big)\Bigg]^{1/2}.
\eqx

Finally, the complete set of equations determining the superconducting state properties is as follows,

\begin{widetext}
\eqn
 \mathcal{F} &=& - k_B T \sum_{\bk \sigma} \ln (1 + e^{-\beta E_{\bk \sigma}}) + \sum_\bk (\xi^{(s)}_\bk - E_\bk) + N \frac{\Delta^2}{V_0} + \mu N + \frac{N}{n} \overline{m} h_{cor}, \label{eq:sc1} \\
 h_{cor} &=& - \frac{n}{N} \sum_{\bk \sigma} f(E_{\bk \sigma}) \frac{\partial E_{\bk \sigma}}{\partial \overline{m}} + \frac{n}{N} \sum_\bk \frac{\partial \xi_\bk^{(s)}}{\partial \overline{m}} \Big( 1 - \frac{\xi_\bk^{(s)}}{E_\bk} \Big), \label{eq:sc2} \\
 \overline{m} &=& \frac{n}{N} \sum_{\bk \sigma} \sigma f(E_{\bk \sigma}), \label{eq:sc3}\\
 \Delta_\bq &=& \frac{V_0}{N} \sum_\bk \frac{1 - f(E_{\bk \uparrow}) - f(E_{\bk \downarrow})}{2 E_\bk} \Delta_\bq, \label{eq:sc4} \\
 n &=& n_\uparrow + n_\downarrow = \frac{n}{N} \sum_{\bk \sigma} \Big\{u_\bk^2 f(E_{\bk \sigma}) + v_\bk^2 \big[1 - f(E_{\bk, -\sigma})\big]\Big\}, \label{eq:sc5}
\eqnx
\end{widetext}

where $\mathcal{F}(T, H_a; h_{cor}, \overline{m}, n, \Delta_\bq)$ is the system free energy for the case of a fixed number of particles \cite{Koponen}. Similarly as for the non-interacting Fermi sea, the equations (\ref{eq:sc2}), (\ref{eq:sc3}) and (\ref{eq:sc4}) are equivalent with $\partial \mathcal{F} / \partial \overline{m} = 0$, $\partial \mathcal{F} / \partial h_{cor} = 0$ and $\partial \mathcal{F} / \partial \Delta_\bq = 0$, respectively. In effect, the numerical analysis involves solving the system of four integral equations. Note also the presence of two different effective chemical potentials $\mu_\sigma = \mu + \sigma h_{cor}$ for particles with spin up and down in the spin-polarized situation. This is an unavoidable consequence (c.f. Refs. \cite{g, h, Spalek2, hmol}) of the slave-boson formalism used to derive expression for the masses (\ref{eq:m}), and dispersion relation (\ref{eq:disp}). Parenthetically, we have also performed calculations by disregarding the different effective chemical potentials (i.e. we have put $h_{cor} = 0$) and the results obtained were nonphysical (the free-energy jump occurred at the BCS-FFLO phase transition). Obviously, the physical chemical potential is $\mu$ and is determined from the neutrality condition (\ref{eq:sc5}).

The physical solution is that with a particular $\bq$ which minimizes the free energy (\ref{eq:sc1}). The state with $\bq = 0$ is called the BCS state, and that with $|\bq| \neq 0$ - the FFLO state.

\section{Numerical Analysis and Discussion}
\label{sec:numerics}

We assumed, the following values of the parameters, emulating the heavy fermion systems: $n = 0.97, \,\, V = 161 \, \AA^3$, $m_{av} = 100 \, m_0$ (data for CeCoIn$_5$), $\hbar \omega_C = 17 \, K$, and $V_0 = 93 \, K$. The characteristic energy scale associated with spin-fluctuations in CeCoIn$_5$ is $T_{sf} = 10 \, K$ \cite{Petrovic} - a value comparable to our $\hbar \omega_C$. For those parameters, the chemical potential was equal to $\mu \approx 140 \, K$. This means that $V_0 \sim \epsilon_F$ and the (weak coupling) BCS approximation can be regarded only as a proper solution on a semiquantitive level at best. Furthermore, for the values of parameters one can calculate the coupling constant $\rho(\mu) V_0 \approx 0.48$ and the coherence length at $T = 0 \, K$, $\xi_0 \approx 40 \AA$, both already at the border of the strong-coupling limit. Also, such values have been taken to obtain the critical temperature $T_s \simeq 2.3 K$.

\begin{figure}
\scalebox{1.2}{\includegraphics{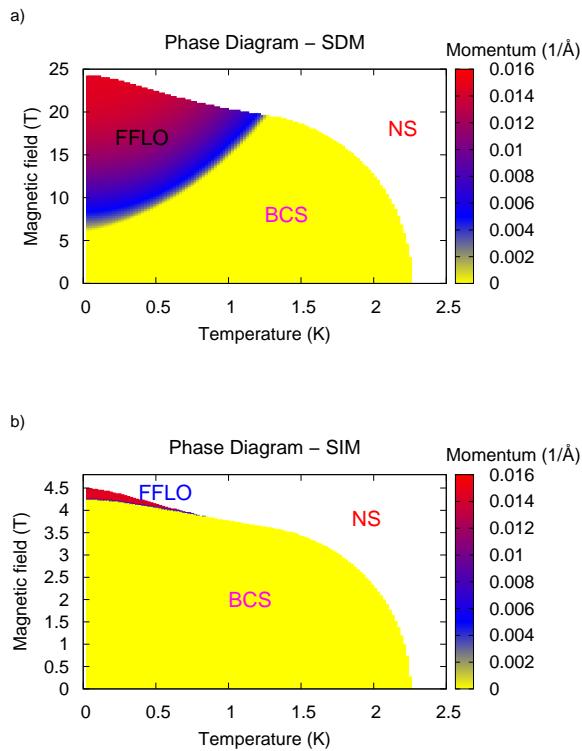}}
\caption{\label{fig:PHD} (Color online) Phase diagram for the cases with the spin-dependent a) and the spin-independent masses b). Light (yellow) region corresponds to $\bq = 0$ (BCS phase), the darker (blue-red) one to $\bq \neq 0$ (FFLO phase) and the white to normal state.
Note that with increasing temperature, the transition from BCS to FFLO state occurs at higher fields, in qualitative agreement with experimental results \cite{CeCo2}. The FFLO phase is stable in an extended $H_a$-$T$ regime only in the SDM case.}
\end{figure}

\begin{figure}
\scalebox{1}{\includegraphics{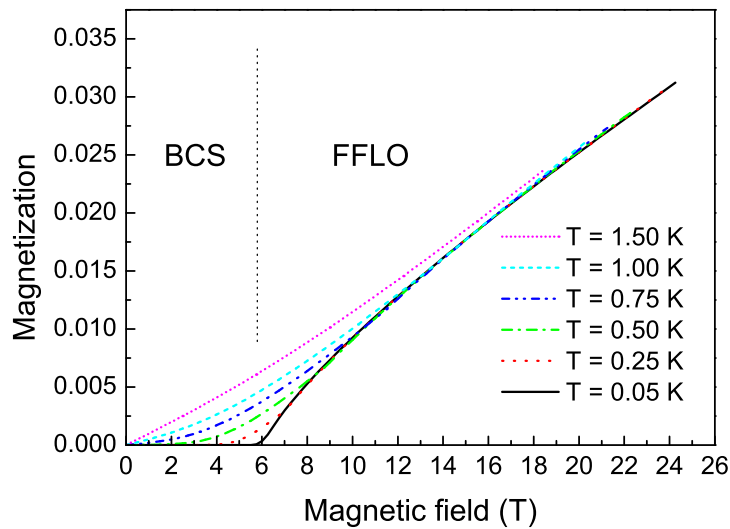}}
\caption{\label{fig:Magnet} (Color online) Spin magnetization as a function of temperature and applied magnetic field. For $T = 0 - 0.5 \, K$, the magnetization in the BCS state is small, than it increases with applied field after the continuous BCS to FFLO transition. For higher temperatures magnetization in the BCS state becomes substantial and this produces a higher critical field for the BCS-FFLO transition for $T \gtrsim 0.5 \, K$.}
\end{figure}

We now discuss the phase diagrams for the cases of SDM and SIM which are exhibited in Figs. \ref{fig:PHD}a-b, respectively. Both the BCS (state with $\bq = 0$) and the FFLO ($\bq \neq 0$) phases extend to much higher fields if the masses are spin-dependent. This is a consequence of the smaller Fermi-vector splitting $\Delta k_F$ for the SDM case (c.f. Fig. \ref{fig:FS}d). The most interesting is the fact that in the SDM situation the FFLO state becomes much more robust compared to BCS state, especially for $T \simeq 0$. The reason for this is as follows: the superconductivity in the Pauli limiting case is destroyed by the Fermi vectors splitting $\Delta k_F$ (c.f. Fig. \ref{fig:FS}d). This splitting in the case of SDM is generally smaller (in this respect SDM compensate effect of the Zeeman term), hence the higher critical fields. However, for the masses to depend on spin, the magnetization $\overline{m} \equiv n_\uparrow - n_\downarrow$ has to be non-zero, and in the BCS state around $T \lesssim 0.5 \, K$ magnetization is close to zero (see Fig. \ref{fig:Magnet}) what weakens the mass dependence on spin and in effect produces larger $\Delta k_F$. Therefore, the BCS state is not enhanced much by SDM in that temperature interval. In the FFLO state, on the other hand the magnetization is nonzero even at $T= 0 \,K$.
This is because in the FFLO state there are regions with unpaired quasiparticles in the reciprocal space. The FFLO state becomes stable in this regime, as a result of a smaller Fermi vectors splitting.

Another interesting feature is the fact that with increasing temperature, the transition from BCS to FFLO state occurs at high fields (cf. Fig. \ref{fig:PHD}a) consistent with experimental results \cite{CeCo2}. It can also be easily explained. As temperature increases, the magnetization in the BCS state increases, allowing a substantial mass difference, and decreasing the Fermi wavevectors splitting, enhancing superconductivity. Therefore, the BCS state benefits from the smaller $\Delta k_F$ for SDM at higher temperatures ($T \gtrsim 0.5 \, K$) and becomes more stable in this regime.

Systematic evolution of the pure spin magnetization in the condensed state is shown in Fig. \ref{fig:Magnet} (the orbital part is not included). It increases at the BCS-FFLO border al lower $T$, as one would expect.

\begin{figure*}
\scalebox{0.9}{\includegraphics[angle=270]{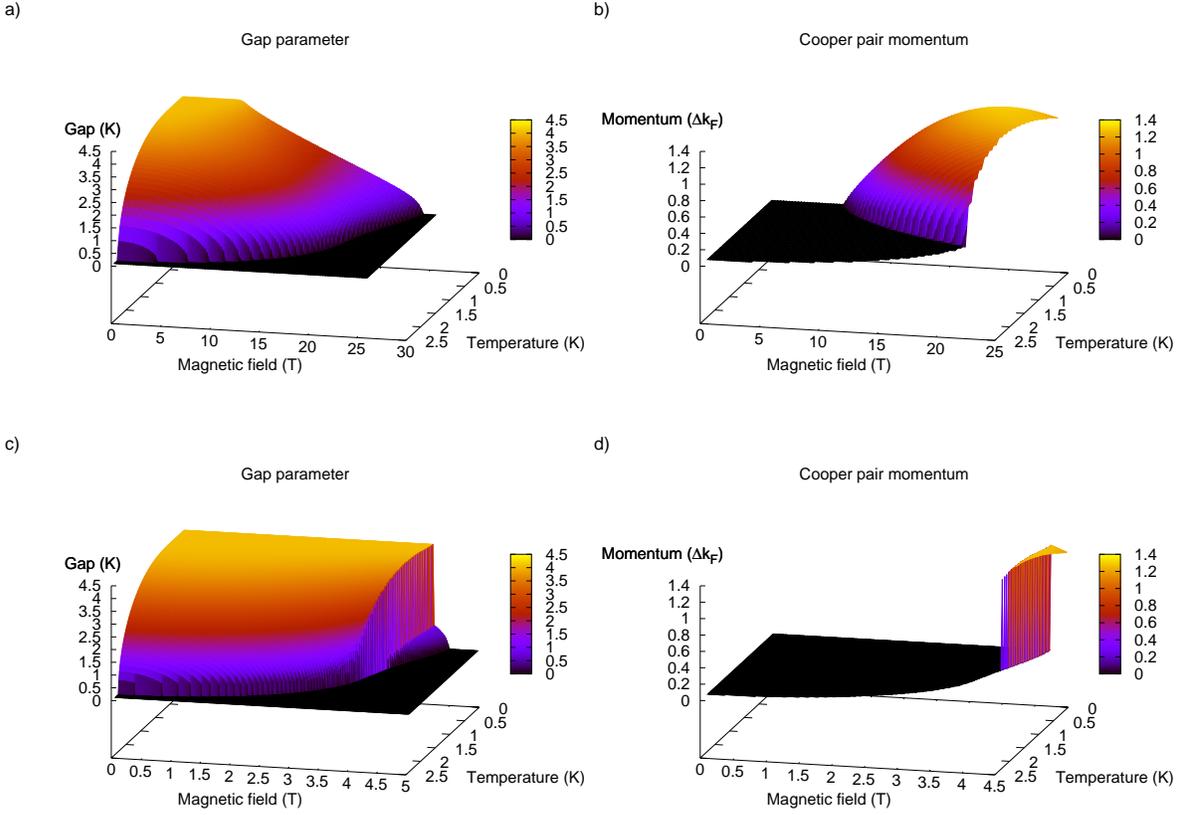}}
\caption{\label{fig:Del} (Color online) Left (a, c): gap parameter $\Delta_\bq$ as a function of temperature and magnetic field for SDM (up) and SIM (down). All transitions for SDM are continuous. Right (b, d): Cooper pair momentum in the FFLO state in units of Fermi-wavevector splitting $\Delta k_F$ for SDM (up) and SIM (down). Note that for the FFLO phase the momentum $|\bq|$ changes continuously (with transition BCS-FFLO), contrary to the case of SIM. Typical value of the momentum is $|\bq| \approx \Delta k_F$. }
\end{figure*}

In the panel composing Fig. \ref{fig:Del} we plot the gap magnitude $\Delta_\bq$ and the magnitude of the wave vector $\bq$ for SDM and SIM cases. The behavior of the order parameter $\Delta_\bq$ differs substantially in these two cases. Namely, there is no jump of $\Delta_\bq$ at BCS-FFLO transition for SDM, whereas for SIM this transition is always discontinuous. Transitions from superconducting to normal state are continuous for the case of SDM in disagreement with the experimental results \cite{CeCo2}. The reasons for this discrepancy are discussed in the next Section.

\begin{figure}
\scalebox{1.2}{\includegraphics{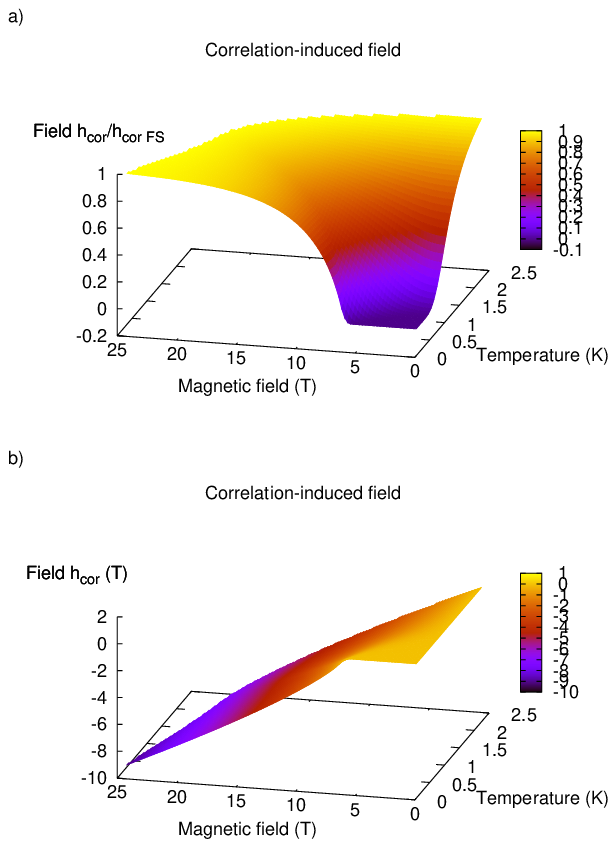}}
\caption{\label{fig:hcor} (Color online) Left: Correlation field in units of the field obtained for non-interacting Fermi sea with SDM included. Note that as $H_a \rightarrow H_{c2}$, the correlation-field value approaches the one for the unpaired Fermi sea. Right: Correlation field in absolute units. It is negative, i.e. it acts opposite to the applied magnetic field.}
\end{figure}

Finally in Fig. \ref{fig:hcor} we show the correlation-field dependence. It can be seen that for BCS around $T=0 \, K$ this field is close to zero, then increases and approaches for $H_a \rightarrow H_{c2}$ the value for the unpaired Fermi sea, denoted here as $h_{cor FS}$.

\section{Outlook and Conclusions}   \label{sec:summary}

We have analyzed superconducting states of a three-dimensional gas of heavy quasiparticles with (SDM) and without (SIM) the spin-dependent masses. Despite the simplicity of our model (parabolic dispersion relation, constant pairing potential, $s$-wave gap, single narrow-band model), qualitative results obtained are very meaningful for the FFLO phase detectability and should hold for more general and realistic models and other high-field low-temperature (HFLT) phases. This is because the spin-dependent factor renormalizing mass is $\bk$-independent, as is $h_{cor}$, and they are obtained in a self-consistent manner from global equations, integrated over $\bk$. In effect, their values should not be influenced strongly by the details of the bare electronic structure.

The most striking result is the fact that for the case of SDM the FFLO state becomes stable in much wider range of applied field and temperature. We believe that the mechanism of stabilization of the FFLO state by SDM is universal. Therefore, it should also apply to other unconventional HFLT phases such as for example the mixed staggered $\pi$-triplet SC + $d$-wave singlet SC + SDW phase proposed very recently \cite{Aperis1}. This is because in such phases the spin-magnetization is always higher than for the conventional BCS ($s$-wave or $d$-wave) state \cite{Aperis1, Mitrovic}. Those phases will benefit, even to a larger extent, from the compensation of Zeeman effect by SDM, on the expense of the BCS phase, as discussed earlier.

The detailed application of our results to concrete systems is rather limited. This is because three topics require still a conjoined analysis from the theoretical side. This is the inclusion of the singlet-triplet mixing in the FFLO phase \cite{Aperis1, ST}, as mentioned in the Cooper-pair case (cf. Sec. \ref{sec:Cooper}). Associated with it is the problem of magnetism appearing in the normal portion of the system in the FFLO state. Third, we have to introduce $d$-wave symmetry of the superconducting gap. Inclusion of those factors introduces additional self-consistent integral equations making the whole approach much more complex from the numerical side. Then, one has also to carry out the whole procedure for a realistic (quasi-two-dimensional) electronic structure. The inclusion of magnetism should result in the first-order nature of the BCS-FFLO phase boundary \cite{Sigrist, Aperis1}. We should be able to see a progress along these lines in the near future.

The nature of the HFLT unconventional phase in the heavy - fermion system CeCoIn$_5$ is still unclear. Some studies suggest FFLO character \cite{Movshovich}, others reject it \cite{Kenzelmann}. We claim that whatever this state really is, it may be stabilized by SDM due to its higher spin-susceptibility \cite{Mitrovic}. So far the observation of FFLO phase in organic metal has been confirmed \cite{Singleton}, but no spin-dependence of the effective mass has been investigated for those systems.

To conclude, the simultaneous observation of the spin-dependent masses and of an unconventional HFLT superconducting phase in the same system should not be regarded as coincidental. Hence, other unconventional HFLT phases may be searched for in the systems in which spin-split masses have been observed and \textit{vice versa}.

\begin{acknowledgments}
The authors are grateful to Jakub J\c{e}drak for interesting discussions and suggestions concerning numerical determination of $h_{cor}$.
The work was supported by Ministry of Higher Education and Science, Grants Nos. N N202 128736 and N N202 173735. The project was performed under the auspices of the COST P-16 Grant of the European Science Foundation, entitled \textit{"Emergent Behavior in Correlated Matter"} (ECOM), as well as of the National Network \textit{"Strongly Correlated Systems"}.
\end{acknowledgments}

\appendix

\section{Elementary derivation of the spin-dependent mass enhancement factor \label{appA}}
The mass enhancement factor in a strongly correlated system can be understood on the intuitive ground in the following manner. Usually, it is the most decisive factor in determining the quasiparticle density of states \cite{g, Bauer, Bauer2}, so the argument is carried out for featureless (rectangular) form of the density of states in the bare band (here taken per site and one spin direction), i.e.

\eq
\rho(\epsilon) = \left\{
\begin{array}{ll}\frac{1}{W} , &\textrm{for } -\frac{W}{2} \leq \epsilon \leq \frac{W}{2}\\
0, & \textrm{otherwise,}
\end{array} \right.
\eqx

where $W$ is the bare bandwidth. In such situation the bare band energy per site is

\eq
\Big(\frac{E_B}{N}\Big)_{U \rightarrow 0} = \int_{-\frac{W}{2}}^{\mu} \epsilon \rho(\epsilon) d\epsilon = -\Big(\frac{W}{2}\Big) \sum_\sigma n_\sigma (1 - n_\sigma),
\label{eq:a2}
\eqx

where the chemical potential is defined from

\eq
n_\sigma = \int_{-\frac{W}{2}}^{\mu} \rho(\epsilon) d\epsilon.
\eqx

On the other hand, for strongly correlated electrons $(U \rightarrow \infty)$ the corresponding energy of itinerant electrons is

\eq
\Big(\frac{E_B}{N}\Big)_{U \rightarrow \infty} = -\Big(\frac{W}{2}\Big) \sum_\sigma n_\sigma (1 - n).
\label{eq:a4}
\eqx

The factor $(1-n)$ expresses the fact that the hopping of the electron between the neighbors takes place when there is no other electron present on neighboring site. Combining (\ref{eq:a2}) and (\ref{eq:a4}) one can write down that

\eq
\Big(\frac{E_B}{N}\Big)_{U \rightarrow \infty} = -\Big(\frac{W}{2}\Big) \sum_\sigma q_\sigma n_\sigma (1 - n_\sigma) = \sum_{k<k_F, \sigma} q_\sigma \epsilon_\bk,
\eqx

with $q_\sigma = (1-n)/(1-n_{\sigma})$ is the spin dependent, $\bk$-independent renormalization factor. We write now the quasiparticle energy in the form $E_{\bk \sigma} = q_\sigma \epsilon_\bk \equiv \epsilon_\bk + (q_\sigma - 1) \epsilon_\bk \equiv \epsilon_\bk + \Sigma_\sigma(\epsilon_\bk)$, where $\Sigma_\sigma(\epsilon_\bk)$ is the self-energy part (its real part) induced by the correlations. Defining the mass via the standard Fermi-liquid relation

\eq
\Big( \frac{m_\sigma}{m_B} \Big)^{-1} = 1 + \frac{\partial\Sigma_\sigma(\omega)}{\partial \omega}|_{\omega = \epsilon_\bk},
\eqx

we obtain that $m^* \equiv m_\sigma = m_B$ and $1/q_\sigma = m_B (1-n_{\sigma})/(1-n)$, which is equivalent to (\ref{eq:m}). The spin-dependent masses of this form appear in either Gutzwiller \cite{g} or slave-boson approaches \cite{Spalek2} which reduce to the Gutzwiller-ansatz results in the paramagnetic case. They do not appear in the simple slave-boson approach \cite{Wassermann}, but then the overall quasiparticle mass depends on applied field. The appearance of the effective field $h_{cor}$ \cite{g, h, Spalek2, hmol} arises from a constraint on the number $n_\sigma$ of fermions, in direct analogy to the introduction of chemical potential in grand-canonical approach.

\section{Pair binding energy for general antisymmetric state\label{appB}}

We consider here the most general wave function antisymmetric with respect to transposition of particles (\ref{eq:trafocz}) and show that its binding energy decreases with the increasing center-of-mass momentum $\bQ$ and is lower than the energy of the specific-spin state (\ref{eq:spin1}). In the main text we refer to this solution as to the singlet (which is a specific case of the considered here general function), because we would like to recover the standard singlet solution in the $H_a = 0$ limit. Spin states with a well-defined transposition symmetry (singlet and triplet with $S^z = 0$) are described by the following wave functions

\begin{eqnarray}
&&\chi_S(\sigma_1, \sigma_2) = \frac{1}{\sqrt{2}} (|1\uparrow\rangle|2\downarrow\rangle -
|1\downarrow\rangle|2\uparrow\rangle), \\
&&\chi_T(\sigma_1, \sigma_2) = \frac{1}{\sqrt{2}} (|1\uparrow\rangle|2\downarrow\rangle +
|1\downarrow\rangle|2\uparrow\rangle).
\end{eqnarray}

Therefore, we can expand the general antisymmetric wave function in the basis of singlet and triplet wave functions.

\begin{equation}
\Psi(\mathbf{r_1}, \mathbf{r_2}, \sigma_1, \sigma_2) = \lambda \, \Phi_S(\mathbf{r_1}, \mathbf{r_2}) \chi_S(\sigma_1, \sigma_2) + \sqrt{1-\lambda^2} \, \Phi_A(\mathbf{r_1}, \mathbf{r_2}) \chi_T(\sigma_1, \sigma_2). \label{eq:funkcjaog}
\end{equation}

Where $\Phi_S$ is symmetric and $\Phi_A$ antisymmetric under ($\mathbf{r_1} \leftrightarrow \mathbf{r_2}$) and $\lambda$ is characterizing the degree of mixing. We underline once again that (\ref{eq:funkcjaog}) is the most general antisymmetric wave function.

We express the functions $\Phi_A$ and $\Phi_S$ as a superposition (we take $V=1$ for simplicity)

\begin{eqnarray}
&& \sqrt{1-\lambda^2} \, \Phi_A(\mathbf{r_1}, \mathbf{r_2}) = \sum_{\mathbf{k_1}, \mathbf{k_2}} \alpha_{\mathbf{k_1}, \mathbf{k_2}} \bigg( e^{i \mathbf{k_1}\mathbf{r_1}+ i\mathbf{k_2}\mathbf{r_2}} - e^{i \mathbf{k_1}\mathbf{r_2}+ i\mathbf{k_2}\mathbf{r_1}}\bigg), \\
&& \lambda \, \Phi_S(\mathbf{r_1}, \mathbf{r_2}) = \sum_{\mathbf{k_1}, \mathbf{k_2}} \beta_{\mathbf{k_1}, \mathbf{k_2}} \bigg( e^{i \mathbf{k_1}\mathbf{r_1}+ i\mathbf{k_2}\mathbf{r_2}} + e^{i \mathbf{k_1}\mathbf{r_2}+ i\mathbf{k_2}\mathbf{r_1}}\bigg).
\end{eqnarray}

Therefore, the wave functions have the proper symmetry by construction. Setting either $\alpha_{\mathbf{k_1}, \mathbf{k_2}} = 0$ or $\beta_{\mathbf{k_1}, \mathbf{k_2}} = 0$, we obtain the singlet or triplet wave functions, respectively. Note that the binding energy for the triplet state vanishes for the $s$-wave pairing. We consider the Cooper problem with the wave function (\ref{eq:funkcjaog}) and Hamiltonian in the form (\ref{eq:HMOP}).

As we already have said, the natural basis of the spin wave functions is spanned by the specific-spin wave functions. For this reason, we transform the wave function (\ref{eq:funkcjaog}) to the form

\begin{eqnarray}
\nonumber  \Psi(\mathbf{r_1}, \mathbf{r_2}, \sigma_1, \sigma_2) = \sum_{\mathbf{k_1}, \mathbf{k_2}} \Big[ &\big( \lambda_{\mathbf{k_1}, \mathbf{k_2}} e^{i \mathbf{k_1}\mathbf{r_1} + i\mathbf{k_2}\mathbf{r_2}} + \gamma_{\mathbf{k_1}, \mathbf{k_2}} e^{i \mathbf{k_1}\mathbf{r_2} + i\mathbf{k_2}\mathbf{r_1}} \big) &|1\uparrow\rangle|2\downarrow\rangle - \\
&\big( \gamma_{\mathbf{k_1}, \mathbf{k_2}} e^{i \mathbf{k_1}\mathbf{r_1} + i\mathbf{k_2}\mathbf{r_2}} + \lambda_{\mathbf{k_1}, \mathbf{k_2}} e^{i \mathbf{k_1}\mathbf{r_2} + i\mathbf{k_2}\mathbf{r_1}} \big) &|1\downarrow\rangle|2\uparrow\rangle\Big],
\end{eqnarray}

where the new coefficients are given by

\begin{eqnarray}
\lambda_{\mathbf{k_1}, \mathbf{k_2}} \equiv \beta_{\mathbf{k_1}, \mathbf{k_2}} + \alpha_{\mathbf{k_1}, \mathbf{k_2}},\\
\gamma_{\mathbf{k_1}, \mathbf{k_2}} \equiv \beta_{\mathbf{k_1}, \mathbf{k_2}} - \alpha_{\mathbf{k_1}, \mathbf{k_2}}.
\end{eqnarray}

Following the standard but cumbersome procedure, we obtain the following two equations for those expansion coefficients

\begin{eqnarray}
\lambda_{\mathbf{k'}}+ \gamma_{-\mathbf{k'}+\bQ\frac{\Delta m}{M}} &=& \frac{1}{N} \frac{\sum_{\mathbf{k}} (\lambda_{\mathbf{k}}+ \gamma_{-\mathbf{k}+\bQ\frac{\Delta m}{M}}) V_{\mathbf{k' k}} }{ \frac{\hbar^2\bQ^2}{2M} + \frac{\hbar^2\mathbf{k'}^2}{2\mu_{red}} - 2 \mu - E }, \label{eq:og1e} \\
\lambda_{\mathbf{k'}} + \gamma_{-\mathbf{k'}+\bQ\frac{\Delta m}{M}} &=& \frac{1}{N} \frac{\sum_{\mathbf{k}} (\lambda_{\mathbf{k}}+ \gamma_{-\mathbf{k}+\bQ\frac{\Delta m}{M}}) V_{\mathbf{(-\mathbf{k'}+\bQ\frac{\Delta m}{M})\,(-\mathbf{k}+\bQ\frac{\Delta m}{M}) }} }{\frac{\hbar^2\bQ^2}{2M} + \frac{\hbar^2\mathbf{k'}^2}{2\mu_{red}} - 2 \mu - E}. \label{eq:og2e}
\end{eqnarray}

For the above equations not to be contradictory, the matrix elements of the pairing potential have to be identical

\begin{equation}
  V_{\mathbf{(-\mathbf{k'}+\bQ\frac{\Delta m}{M})\,(-\mathbf{k}+\bQ\frac{\Delta m}{M}) }} = V_{\mathbf{k' k}}. \label{warpot}
\end{equation}

If this condition is fulfilled, Eqs. (\ref{eq:og1e}) and (\ref{eq:og2e}) are equivalent. We introduce the potential in the standard form with a constant attraction

\begin{equation}
V_{\mathbf{k k'}} = \left\{
\begin{array}{ll} -V_0, &\textrm{for } \mathbf{k}, \mathbf{k'} \in W^*, \\
0, &\textrm{for } \mathbf{k} \notin W^* \vee \mathbf{k'} \notin W^*.
\end{array} \right.
\end{equation}

Now, the condition (\ref{warpot}) is fulfilled if and only if $W^*$ has the following property

\begin{equation}
  W^* = -W^* + \bQ\frac{\Delta m}{M}.
\end{equation}

Clearly, the region $\overline{W}_\bQ$ is not a good choice for $W^*$, but by making use of it we can construct a proper interaction region as follows

\begin{equation}
  W^* = \overline{W}_\bQ \cap (- \overline{W}_\bQ + \bQ\frac{\Delta m}{M}).
\end{equation}

Such interaction region has a physical meaning, as we can rewrite it in the form

\begin{eqnarray}
&W^* =& \{ \mathbf{k} = \frac{\mathbf{k_1} m_2 - \mathbf{k_2} m_1}{m_1 + m_2}
\,|\, \mathbf{k_1} \in W_1 \cap W_2 \wedge \mathbf{k_2} \in W_2 \cap W_1 \wedge \bQ = \mathbf{k_1} + \mathbf{k_2} \}, \\
&W_i =& \{ \mathbf{k} \,|\, 0 \leq \xi_{\bk\sigma_i} \leq \hbar \omega_C \} \textrm{, } i=1,2,
\end{eqnarray}

meaning that the interaction takes place in the regions of $\bk$-space, for which not only both particles are at most $\hbar \omega_C$ Fermi level, but also if we exchanged their wavevectors ($\bk_1 \leftrightarrow \bk_2$), they are still at most $\hbar \omega_C$ above their Fermi surface.

The equation for the binding energy can be obtained from (\ref{eq:og1e}) after summing up over $\mathbf{k'} \in W^*$ and dividing by $\sum_{\mathbf{k} \in W^*} (\lambda_{\mathbf{k}}+ \gamma_{-\mathbf{k}+\bQ\frac{\Delta m}{M}})$. In effect, we have that

\begin{equation}
\frac{N}{V_0} = \sum_{\mathbf{k} \in W^*} \frac{1}{ \frac{\hbar^2\bQ^2}{2M} + \frac{\hbar^2\mathbf{k}^2}{2\mu_{red}} - 2 \mu - E}.\label{eq:oggap1}
\end{equation}

This equation has the same form as the one obtained for the \textit{specific-spin} state, the only difference being the appearance of the interaction region $W^*$ instead of $\overline{W}_\bQ$. From this very fact follows automatically the lower binding energy for the antisymmetric state (\ref{eq:funkcjaog}), as $W^* \subset \overline{W}_\bQ$. For the case of pair at rest ($\bQ = 0$), both states (specific-spin and antisymmetric) have equal binding energy, because then $W^* = \overline{W}_\bQ = 0$. The numerical results show that the binding energy of the antisymmetric state decreases linearly with the increasing $\bQ$ (see Fig. \ref{fig:GBQ}).

\section{Narrow-band limit of the Anderson lattice for U $\rightarrow \infty$ \label{appC}}

In this paper we consider the narrow-band limit of the Kondo-lattice state. Such limit has been considered earlier \cite{2s}; here we provide a simple analytic argument. For the sake of simplicity we consider only $h = 0$ case, but include also the hybrid pairing due to the Kondo-type coupling in the strong-correlation limit. The starting Hamiltonian with the term $\sim V^2/U$ included, has the form

\eq
\mathcal{\tilde{H}} = \sum_{\langle m, n \rangle \sigma} t_{m, n} c^\dagger_{m \sigma} c_{n \sigma} + \epsilon_f \sum_{i \sigma} \tilde{N}_{i \sigma} + V \sum_{i \sigma}\Big( \tilde{f}^\dagger_{i\sigma} c_{i \sigma} + c^\dagger_{i \sigma} \tilde{f}_{i \sigma}\Big) - \frac{2 V^2}{\epsilon_f + U} \sum_{i\sigma} \tilde{b}^\dagger_{ii}\tilde{b}_{ii},
\eqx

where the first term represents the band energy of conduction electrons, the second the atomic energy of $f$ electrons ($\tilde{N}_{i\sigma} = \tilde{f}^\dagger_{i\sigma} \tilde{f}_{i\sigma} $), the third is the hybridization part, and the last the local pairing with $\tilde{b}^\dagger_{ii} = (1/\sqrt{2}) (\tilde{f}^\dagger_{i\uparrow} c^\dagger_{i \downarrow} - \tilde{f}^\dagger_{i \downarrow} c^\dagger_{i \uparrow})$. The tilted operators exclude the double occupancies of $f$ states, i.e. $\tilde{f}_{i\sigma} \equiv f_{i\sigma} (1-f^\dagger_{i \overline{\sigma}} f_{i \overline{\sigma}})$ and $\tilde{f}^\dagger_{i\sigma} \equiv f^\dagger_{i\sigma} (1-f^\dagger_{i \overline{\sigma}} f_{i \overline{\sigma}})$, $U$ is the magnitude of the Hubbard $f-f$ interaction. In the slave-boson saddle-point approximation this Hamiltonian representing the states in the lowest hybridized band reduces to

\eq
\mathcal{\tilde{H}} = \sum_{\bk \sigma} E_{\bk -} \alpha^\dagger_{\bk \sigma} \alpha_{\bk \sigma} - \frac{4 V^2}{\epsilon_f + U} \sum_{\bk \bk'} \frac{ \tilde{V}^2}{[(\epsilon_\bk - \tilde{\epsilon}_f)^2 + 4\tilde{V}^2 ]^{1/2}[(\epsilon_{\bk'} - \tilde{\epsilon}_f)^2 + 4\tilde{V}^2 ]^{1/2}} \alpha^\dagger_{\bk \uparrow} \alpha^\dagger_{\bk \downarrow} \alpha_{-\bk' \downarrow} \alpha_{\bk' \uparrow}, \label{eqC:Ham}
\eqx

where

\eq
E_{\bk - } = \frac{1}{2} \Big\{ \epsilon_\bk + \tilde{\epsilon_f} - \big[ (\epsilon_\bk - \tilde{\epsilon}_f)^2 + 4 \tilde{V}^2\big]^{1/2} \Big\}
\eqx

and the $\tilde{\epsilon}_f$ and $\tilde{V}$ are the renormalized quantities $\epsilon_f$ and $V$ respectively.

Taking in the pairing part the $\bk$ states close to the bare Fermi energy, we have that $\epsilon_\bk - \tilde{\epsilon}_f \sim - \tilde{\epsilon}_f \sim \tilde{V}$ and under these circumstances the Hamiltonian (\ref{eqC:Ham}) reduces to the BCS form with $\alpha_{\bk \sigma} \approx f_{\bk \sigma}$ and a weakly $\bk$-dependent pairing potential, which is approximated by a constant $V_0$ in main text (cf. Eq. (\ref{eq:HStart})). This approximation is justified also because the maximum pairing amplitude is achieved for $\epsilon_\bk = \tilde{\epsilon}_f$, when the $\bk$-dependent ratio reduces to $4 \tilde{V}^2$. The detailed estimate of $E_{\bk-}$ and of the pairing potential in terms of the effective Kondo temperature is more subtle \cite{2s}. Roughly, the bandwidth of the heavy-quasiparticle band (of $f$-electrons) can be estimated as due to $f-f$ hopping between the sites $\langle i j \rangle$ and is $(\tilde{V}/\epsilon_f)^2 t_{ij} = q (V/\epsilon_f)^2 t_{ij}$ (since $qV^2 = \tilde{V}^2$). This is because it represents the three-step hopping sequence between neighboring $f$-states via conduction $c$-states: $f \rightarrow c$ transition followed by a hopping in $c$-band and a subsequent $c \rightarrow f$ deexcitation. One should underline that such simple form of the pairing part appears only if the main contribution to the hybridization $V = V_{im}$ is of intraatomic ($i=m$) character.

{}

\end{document}